\documentclass[a4paper,fleqn,usenatbib]{mnras}

\usepackage{newtxtext,newtxmath}

\usepackage[T1]{fontenc}
\usepackage{ae,aecompl}

\usepackage{graphicx}	
\usepackage{amsmath}	
\usepackage{amssymb}	

\title[The variance of DM of high-z transient objects]{The variance of dispersion measure of high-redshift transient objects as a probe of ionized bubble size during reionization}

\author[S. Yoshiura et al.]{
 Shintaro Yoshiura$^1$\thanks{E-mail: 161d9002@st.kumamoto-u.ac.jp},
 Keitaro Takahashi$^1$
\\
$^{1}$Department of Physics, Kumamoto University, Kumamoto,Japan\\
}

\date{Accepted XXX. Received YYY; in original form ZZZ}

\pubyear{2017}

\begin{document}
\label{firstpage}
\pagerange{\pageref{firstpage}--\pageref{lastpage}}
\maketitle

\begin{abstract}
The dispersion measure (DM) of high-redshift $(z \gtrsim 6)$ transient objects such as Fast Radio Bursts can be a powerful tool to probe the intergalactic medium during the Epoch of Reionization. In this paper, we study the variance of the DMs of objects with the same redshift as a potential probe of the size distribution of ionized bubbles. We calculate the DM variance with a simple model with randomly-distributed spherical bubbles. It is found that the DM variance reflects the characteristics of the probability distribution of the bubble size. We find the variance can be measured precisely enough to obtain the information on the typical size with a few hundred sources at a single redshift.
\end{abstract}

\begin{keywords}
cosmology: dark ages, reionization, first stars 
\end{keywords}



\section{Introduction} \label{sec:intro}

The Epoch of Reionization (EoR) is a cosmological transition where intergalactic medium (IGM) was ionized by ionizing photons from astronomical objects in the early universe. This epoch has been observationally studied by the spectra of high-$z$ quasars \citep{2006AJ....132..117F, 2015MNRAS.447..499M} and the CMB optical depth \citep{2016A&A...596A.108P}, which indicate the reionization was completed at $z \sim 6$ and the duration was $\Delta z < 2.7$. However, very little is known about the evolution of ionized fraction and the sources of ionizing photons.

Although first galaxies are a candidate of the ionizing source, their properties are highly uncertain. Especially, the escape fraction of ionizing photons is a key characteristic. The measured escape fraction at low redshifts ranges from 0.01 to 0.1 \citep{Steidel01,Iwata09} and the predictions from numerical simulations are not necessarily consistent with differences of two orders of magnitude. Besides, star formation in galaxies, especially in low-mass galaxies, can be suppressed by feedback processes such as the photo-evaporation and photo-heating \citep{2008MNRAS.390..920O, 2013MNRAS.428..154H}, so that only heavy galaxies may have contributed to the reionization.

The topology of ionized bubbles is sensitive to the property of ionizing sources. One of powerful tools to study ionized bubbles observationally is the 21$\,$cm line due to hyperfine structure of hydrogen atom and the brightness temperature of the 21$\,$cm-line signal reflects the state of IGM including the density of neutral hydrogen and spin temperature. For instance, the statistical quantities such as the power spectrum and bispectrum of the 21cm-line signal can distinguish the model of the ionizing sources \citep{2007MNRAS.376.1680P,2015MNRAS.451..467S,2017MNRAS.468.1542S}. Beside, the topology of ionized bubbles can be directly probed by Minkowski functionals \citep{2006MNRAS.370.1329G,2011MNRAS.413.1353F,2014JKAS...47...49H,2017MNRAS.465..394Y}. However, the detection of 21$\,$cm-line signal is highly challenging due to the presence of huge foreground.

The dispersion measure (DM) is yet another probe of the IGM during the EoR. It is defined as a line-of-sight (LOS) integration of the electron number density. An electromagnetic wave propagating through a plasma is delayed, depending on the frequency $\nu$, with the delay time proportional to $\nu^{-2} \rm DM$. Thus, if we can measure the DM of high-redshift transient objects, the evolution of the electron density during the EoR can be studied. In fact, previous theoretical works have reported that the DM of high-redshift Gamma Ray Bursts (GRBs) is useful to investigate reionization scenarios \citep{2003ApJ...598L..79I,2004MNRAS.348..999I}. 

Another type of transients for the DM measurement is fast radio bursts (FRBs), which have been discovered relatively recently \citep{2007Sci...318..777L,2011MNRAS.415.3065K,2012MNRAS.425L..71K,2013Sci...341...53T,2014ApJ...792...19B,2014ApJ...790..101S,2015Natur.528..523M,2015MNRAS.447..246P,2015ApJ...799L...5R,2016Natur.531..202S,2017ApJ...841L..12B,2017MNRAS.468.3746C,2017MNRAS.469.4465P}. As can be seen in an FRB catalog in \cite{2016MNRAS.460L..30C}, FRBs generally have very large DMs, $375 < {\rm DM} < 1629\,\rm pc~cm^{-3}$. Because Galactic transients typically have DMs less than 250$\,{\rm pc\,cm}^{-3}$, FRBs are considered to be extra-galactic events. Actually, a candidate of the host galaxy of an FRB was reported to be at $z=0.492$ $\pm$ 0.008 \citep{2016Natur.530..453K} although this is still under debate \citep{2017MNRAS.465.2143J}. Very recently, \cite{2017arXiv170101100T} reported the host galaxy of a repeating FRB at $z \sim 0.2$. 


There have been several studies on the potential of FRBs as a new cosmological probe of, for example, dark-energy \citep{2014PhRvD..89j7303Z} and cosmological parameters \citep{2016ApJ...830L..31Y} via the DM. Further, \cite{2016ApJ...824..105A} found the DM and rotation measure contain information on the intergalactic magnetic field in filaments. \cite{2014ApJ...780L..33M} argued that the probability distribution of DMs reflects the amount of missing baryon surrounding halos and showed that the variance of DMs evolves with $z$.

In addition to these low-$z$ studies, high-$z$ FRBs have also been considered. For example, \cite{2014ApJ...797...71Z} showed that FRBs with $z \sim 3$ can be a tool to study the HeII reionization which causes the change in the differential DM, if thousands of FRBs are observed. In \cite{2016JCAP...05..004F}, they found a relation between DM and the optical depth of the CMB suggested that the optical depth can be probed by the measurement of FRBs during EoR. If an FRB is located at $z>6$, the DM is larger than a few thousands $\rm pc~cm^{-3}$. Although FRBs with such a high DM have not been observed, the observability of high-$z$ FRBs was discussed \citep{2016JCAP...05..004F,2017arXiv170606582F}. According to their estimation, the Square Kilometer Array\footnote{\url{https://www.skatelescope.org/}} phase 2 (SKA2) can detect FRBs even at $z>10$. Further, even the SKA phase 1 (SKA1) is expected to be able to observe bright FRBs at high redshifts \cite{2007Sci...318..777L}.

Previous works which studied the DM as a probe of reionization considered the contribution of homogeneous IGM electrons to the DM \citep{2003ApJ...598L..79I,2004MNRAS.348..999I}. As mentioned above, the high detection rate of FRBs is predicted based on the number density of observed FRBs (e.g. \cite{2016MNRAS.460L..30C}) and large surveys by the SKA promise the enormous detection of FRBs from EoR \citep{2016JCAP...05..004F}. This will give us a chance to study higher order statistics of the DM probability distribution and the inhomogeneity of IGM even at high-$z$. Thus, in this work, we study the effect of inhomogeneity of the IGM on the DM, considering the ionized bubbles, and introduce the variance of DMs. We discuss the potential of the DM variance to obtain information on the size distribution of ionized bubbles and estimate the necessary number of samples.

This paper is organized as follows. We introduce the dispersion measure and its variance in section~\ref{sec:DM} and our spherical bubble model in section~\ref{sec:SBM}. Then our primary results on the evolution of variance of dispersion measure are presented in section~\ref{sec:RD} and discuss the detectability in section~\ref{sec:Det}. Finally, we conclude this paper. Through this paper we assume the $\Lambda$CDM cosmology with parameter ($\Omega_{\rm m}, \,\Omega_{\Lambda}, \,\Omega_{\rm b},\, n_{\rm s},\, \sigma_8, H_0$) = (0.31,\, 0.69, \,0.048, \,0.97,\, 0.82, $\rm 68\,km \,s^{-1}\,Mpc^{-1}$) \citep{2015arXiv150201589P}.

\section{Dispersion Measure}\label{sec:DM}

The velocity of an electromagnetic wave which propagates in an ionized plasma is given as $v = c \sqrt{1-\nu_{\rm p}^2/\nu^2}$. Here, $\nu$ is the frequency, $c$ is the speed of light and $\nu^2_{\rm p}=n_{\rm e}e^2/\pi m_{\rm e}$ is the plasma frequency. Then the arrival time of the wave is delayed, compared to the case without the ionized plasma, as,
\begin{eqnarray}
\Delta t = \frac{e^2}{2\pi m_{\rm e} c}\frac{1}{\nu^2}\rm DM.
\label{dt}
\end{eqnarray}
Here the DM is calculated as 
\begin{eqnarray}
{\rm DM}(z_{\rm s}) = \int_0^{z_{\rm S}} \frac{n_{\rm e}(z)}{1+z}dl
\label{eq:DM}
\end{eqnarray}
,where $n_{\rm e}$ is the number density of free electron, the $z_{\rm S}$ is the redshift of the transient object, $dl = -cdz/(1+z) H(z)$ and $H(z) = H_0[\Omega_{\rm m}(1+z)^3 + \Omega_{\Lambda}]^{1/2}$. Hereafter, we ignore the He reionization.

Now we consider observation of $N$ DMs at a single redshift with various LOSs. The DM depends on the electron-density profile along the LOS, which reflects not only the evolution of average neutral fraction but its fluctuations. Previous works considered only average neutral fraction assuming a homogeneous IGM and studied the mean value of the DMs, $\overline{\rm DM}(z) = \frac{1}{N}\sum_{\rm i}^N {\rm DM}_{\rm i}$, as a function of redshift. In this work, we take the effect of the ionized bubble distribution into account. The electron-density profile is different for different LOSs because of the fluctuations in electron density, especially the presence of ionized bubbles. The variance of observed DMs can be estimated as, $\sigma_{\rm DM}^2(z) = \frac{1}{\rm N-1} \sum_{\rm i}^{\rm N} ({\rm DM}_{\rm i}-\overline{\rm DM})^2$. It should be noted that the mean value $\overline{\rm DM}$ is completely determined from the average neutral fraction and not affected by its inhomogeneity.

The variance of the DMs can be calculated from an integration of power spectrum of electron density as \citep{2014ApJ...780L..33M},
\begin{eqnarray}
\sigma_{\rm DM}^2(z')\approx \int_0^{z'}\frac{cdz}{H(z)}(1+z)^{-4}{n}_{\rm e}^2(z)\int \frac{d^2k_{\perp}}{(2\pi)^2}P_{\rm k}(k_{\perp},z),
\label{sigmaDM}
\end{eqnarray}
where $P_{\rm k}(k_{\perp},z)$ is the 3-dimensional spatial power spectrum of the electron number density. The power spectrum is defined by $(2\pi)^3\delta_{D}({\bf{k}}+{\bf{k}}')P_{\rm k}({\bf{k}}) = \langle\tilde{\delta_{x}}({\bf{k}})\tilde{\delta_{x}}({\bf{k}}')\rangle$, where the fluctuation of $x_{\rm HII}$ is written as $\delta_{x}=x_{\rm HII}/\overline{x}_{\rm HII}-1$, tildes denote the Fourier dual and $\langle \rangle$ indicates an ensemble average. From this expression, we see the contribution of free electrons during the EoR to the DM variance can be written as, $\sigma_{\rm DM,EoR}^2(z) = \sigma_{\rm DM}^2(z)-\sigma_{\rm DM}^2(z_{\rm H})$. Here, $z_{\rm H}$ is the redshift of the reionization completion and we set $z = z_{\rm H} = 6.0$, indicated from Lyman-$\alpha$ absorption lines imprinted on quasars spectra.

\section{Spherical bubble model}\label{sec:SBM}

Although numerical simulations or semi-numerical simulations (e.g. 21cmFAST \citep{2011MNRAS.411..955M}) can simulate the process of reionization and distribution of ionized bubbles, the results are often complicated and difficult to interpret. Therefore, in this work, we employ a relatively simple model of reionization taking qualitative features of (semi-)numerical simulations into account, in order to study the effect of IGM inhomogeneities on $\sigma_{\rm DM}$ and give intuitive interpretation on the results. We construct ionized bubble distribution by placing ionized bubbles spatially randomly with a probability distribution of bubble size which will be described later. The number of bubbles are tuned so that the volume-averaged neutral fraction, $\overline{x}_{\rm HII}$, matches any specific value.

\begin{figure}
\centering
\includegraphics[width=6.5cm]{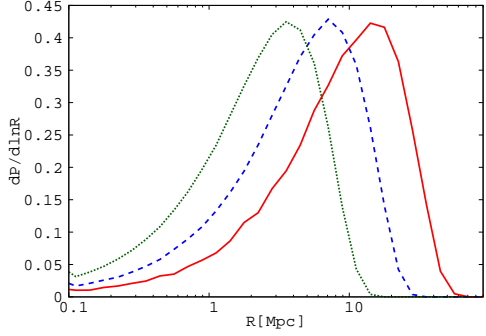}
\caption{Probability density function of ionized bubble size for the model with $\overline{x}_{\rm HII}$=0.5. The dashed line is our fiducial model. The solid and dotted lines are models with $R_{\rm c} = 2 R_{\rm c0}$ and $R_{\rm c0}/2$, respectively. }
\label{fig:hist} 
\end{figure}

An important ingredient of our model is the bubble-size distribution (hereafter BSD). It has been studied both analytically (e.g. \cite{2004ApJ...613....1F}) and numerically (e.g. \cite{2006MNRAS.369.1625I,2007ApJ...654...12Z,2017arXiv170202520K,2017arXiv170600665G}). These BSDs are approximately consistent with each other and can be expressed by a skewed log-normal distribution. Thus, we adopt a skewed log-normal BSD with the mean comoving radius $R_{\rm c}$, standard deviation $\sigma_{\rm L}$ and skewness $\gamma$. It is expected that $R_{\rm c}$ increases, that is, ionized bubbles expand as reionization proceeds. Following \cite{2006MNRAS.365..115F}, we take a fiducial model of $R_{\rm c}$ as,
\begin{equation}
\log R_{\rm c0}(\overline{x}_{\rm HII}) = 2.8 \overline{x}_{\rm HII} - 0.8.
\end{equation}
On the other hand, $\sigma_{\rm L}$ and $\gamma$ are assumed to be constant in time as is implied by previous works and take $\sigma_{\rm L0}^2 = 0.3$ and $\gamma_0 = -1.7$. Fig~\ref{fig:hist} represents the examples of BSD for $\overline{x}_{\rm HII} = 0.5$.

We do not consider partially ionized regions and assume that $x_{\rm HII}$ is unity inside ionized bubbles and $0$ outside. This approximation is reasonable for a reionization process caused by UV photons from galaxies, because UV photons have a short mean free path and are absorbed locally to create ionized bubbles around the sources. Thus, our model would not be valid if reionization is induced mostly by X-rays, which have long mean free path. Further, we do not consider the fluctuations in the electron density within a bubble. This is a good approximation because the fluctuations in electron density are dominated by those in ionized fraction, rather than those in gas density \citep{2015MNRAS.451..467S}

\begin{figure}
\centering
\includegraphics[width=9cm,clip]{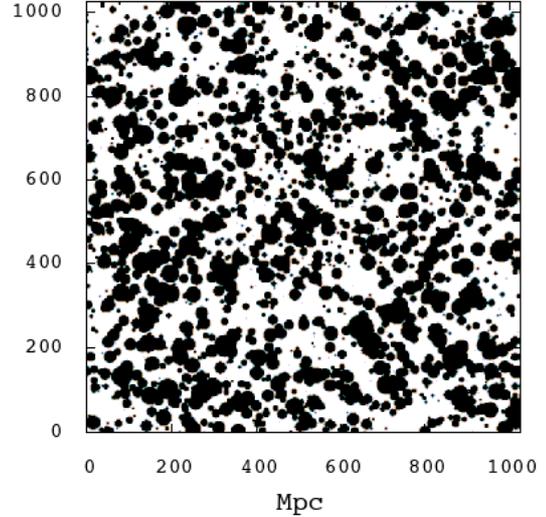}
\caption{Slice of ionized map for fiducial model. The black represents ionized region and the length of a slice is 1024~cMpc. The ionized fraction is $\overline{x}_{\rm HII}$=0.5.}
\label{fig:Maps} 
\end{figure}

With this model, we create free-electron density distribution of a $(1024 \,{\rm cMpc})^3$ box with a resolution of $512^3$. This is done at each redshift from $z_{\rm H}$ to 12 with a step of $\Delta z = 1$, which allows us to perform the integration of Eq.~(\ref{eq:DM}). These boxes with different redshifts are created independently with different seeds of random numbers. This is more reasonable compared to following the evolution of a single box in order to study the free-electron density along a light cone from a transient object. Fig.~\ref{fig:Maps} represents a slice of a box for with $\overline{x}_{\rm HII} = 0.5$. We consider two other models, varying $R_{\rm c}$ as $R_{\rm c} = 2 R_{\rm c0}$ and $R_{\rm c} = R_{\rm c0}/2$, respectively, for a fixed value of $\overline{x}_{\rm HII}$, to see the effect of changing the typical bubble size on $\sigma_{\rm DM}$. For a fixed value of $\overline{x}_{\rm HII}$, a small (large) $R_{\rm c}$ leads to a large (small) number of ionized bubbles. Physically, a model with a small $R_{\rm c}$ and a large number of bubbles corresponds to a case where small (low-emissivity) galaxies contribute to reionization significantly (e.g. \cite{2016MNRAS.459.2342M}). Because the abundance of small galaxies is determined by the effectiveness of the feedback process such as photo-evaporation and photo-heating, the parameter $R_{\rm c}$ effectively parametrizes the feedback efficiency.

\section{Evolution of variance of dispersion measure}\label{sec:RD}

\begin{figure}
\centering
\includegraphics[width=6.5cm]{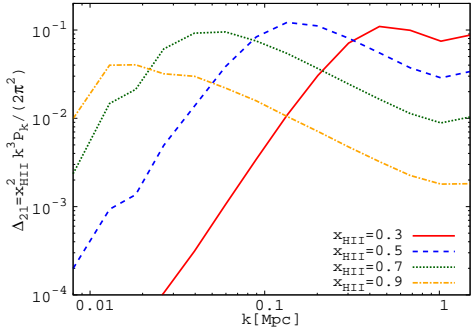}
\caption{The power spectra for the fiducial model with several values of $\overline{x}_{\rm HII}$. Solid, dashed, dotted and dot-dashed lines correspond to $\overline{x}_{\rm HII}=$ 0.3, 0.5, 0.7 and 0.9, respectively. The mean size of ionized bubbles is $R_{\rm c} = $1.1, 4.0, 14 and 52~Mpc, respectively, while the maximum size is 10.4, 36.7, 111 and 313~Mpc, respectively.}
\label{fig:Pk1}
\end{figure}

First, from the above simulations, we can calculate the power spectrum of the fluctuation of the free-electron density. The power spectra with $\overline{x}_{\rm HII} =$ 0.3, 0.5, 0.7 and 0.9 are plotted in the Fig.~\ref{fig:Pk1} for the fiducial model. The power spectra have a peak and the peak wavenumber apparently corresponds to the maximum bubble size (10.4, 36.7, 111 and 313~Mpc, respectively), for these specific realizations) rather than the mean size (1.1, 4.0, 14 and 52~Mpc, respectively). The bubble expansion is clearly seen in the shift of the peak. In spite of its simplicity, the power spectra calculated from our model reasonably agree with those from more sophisticated models (e.g. \cite{2011MNRAS.414..727Z}).

\begin{figure}
\centering
\includegraphics[width=6.5cm]{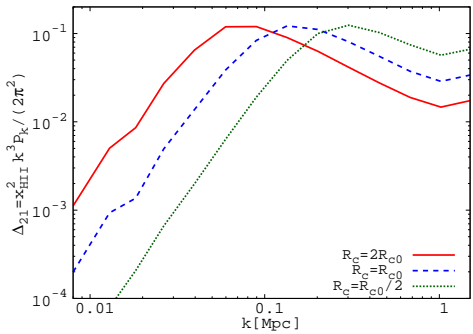}
\caption{The power spectrum for models with $\overline{x}_{\rm HII}$=0.5. Solid, dashed and dotted lines have $2R_{\rm c0}$, $R_{\rm c0}$ and $R_{\rm c0}/2$, respectively.}
\label{fig:Pk2} 
\end{figure}

Fig.~\ref{fig:Pk2} shows the power spectra with $\overline{x}_{\rm HII} = 0.5$ with models of different $R_{\rm c}$. The peak wavenumber shifts according to the value of $R_{\rm c}$. Interestingly, the amplitudes of $P_{\rm k}$ are almost the same for these models and, therefore, the variances of the fluctuations in electron number density, which can be calculated as $\int d^3k P_{\rm k}(k)$, are also the same. 

\begin{figure}
\centering
\includegraphics[width=6.5cm]{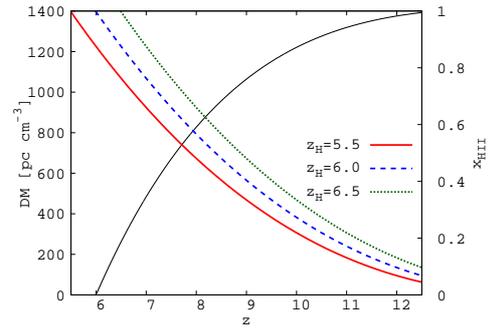}
\caption{Thin solid line represents the redshift evolution of average dispersion measure for our fiducial ionization history ($z_{\rm H} = 6.0$) [left vertical axis]. Other lines show the evolution of $x_{\rm HII}$ for $z_{\rm H} =$ 5.5, 6.0 and 6.5 [right vertical axis].} 
\label{DM} 
\end{figure}

\begin{figure}
\centering
\includegraphics[width=6.5cm]{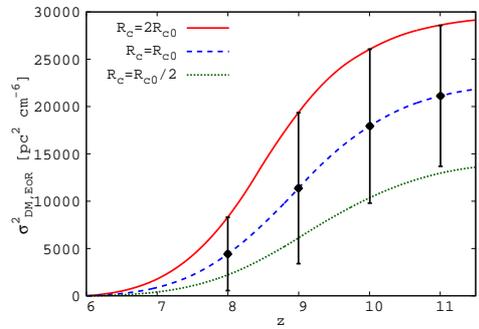}
\caption{The redshift evolution of $\sigma^2_{\rm DM, EoR}$ for the models with $2R_{\rm c}$, $R_{\rm c}$ and  $R_{\rm c}/2$. The vertical lines show the errors with the number of samples listed in Table~1 at $z=$ 8, 9, 10 and 11. } 
\label{sigma} 
\end{figure}

Next, we compare the evolution of $\sigma_{\rm DM,EoR}$ for the three models with different $R_{\rm c}$, fixing the reionization history. The fiducial reionization history with $z_{\rm H} = 6.0$ is shown as the dashed line in Fig.~\ref{DM}, which satisfies the observational constraints from the CMB optical depth to Thomson scattering. Since the three models have the same ionization history, they have the same value of average DM, which is shown as the thin solid line in Fig.~\ref{DM}. In Fig.~ \ref{sigma}, we show the evolution of $\sigma^2_{\rm DM,EoR}$ for the three models as a function of $z$. We find that, as $R_{\rm c}$ increases, $\sigma_{\rm DM,EoR}^2$ evolves more quickly. As we saw in Eq.~\ref{sigmaDM}, $\sigma^2_{\rm DM,EoR}$ is obtained by a 2D integration of the 3D power spectrum. Therefore, the power spectrum which has a peak at a larger scale contributes to $\sigma^2_{\rm DM}$ more significantly. Physically, this is understood as follows. Because a larger value of $R_{\rm c}$ leads to a sparse distribution of large bubbles, the number of bubbles which a light-ray passes is small and has a relatively large statistical fluctuations. In this case, the value of the DM strongly depends on the direction of the LOS, which leads to a large value of $\sigma^2_{\rm DM}$. On the other hand, for a smaller value of $R_{\rm c}$, small bubbles are abundantly distributed so that the statistical fluctuation becomes smaller and a small value of $\sigma^2_{\rm DM}$ is resulted. 

\begin{figure}
\centering
\includegraphics[width=6.5cm]{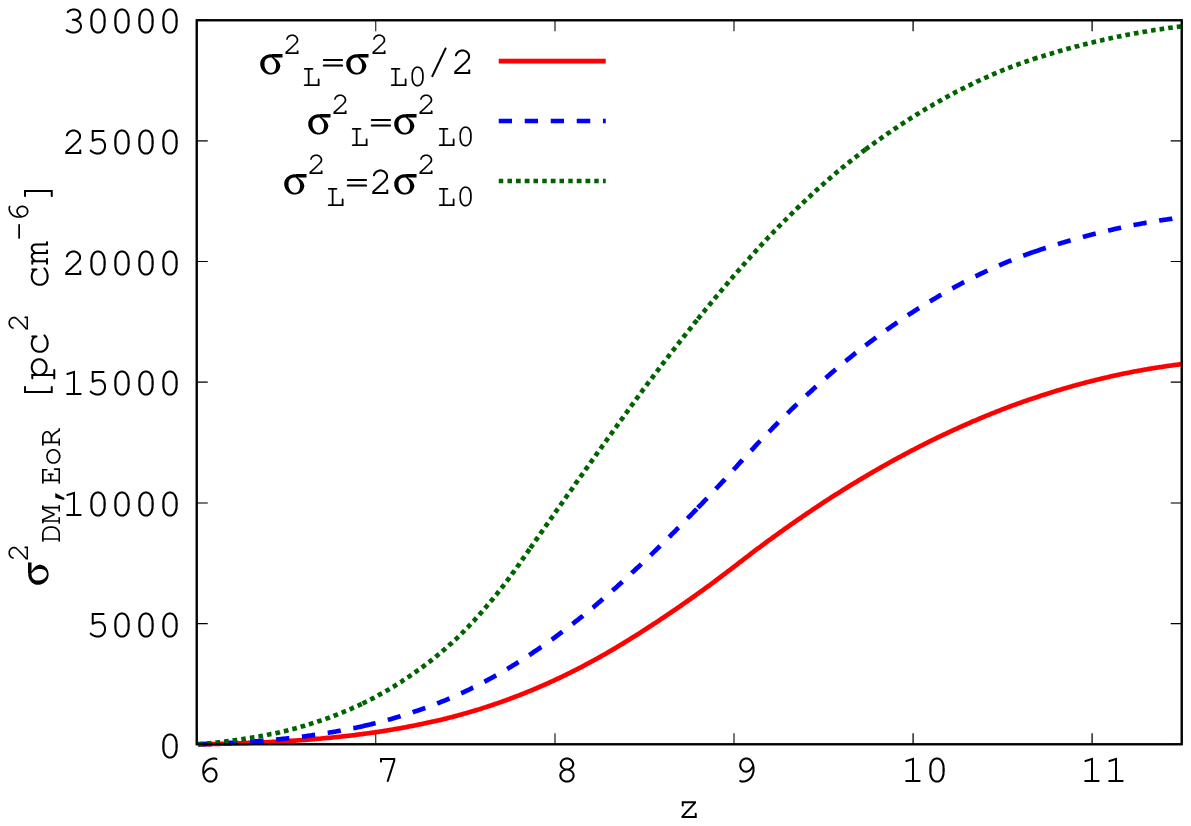}
\includegraphics[width=6.5cm]{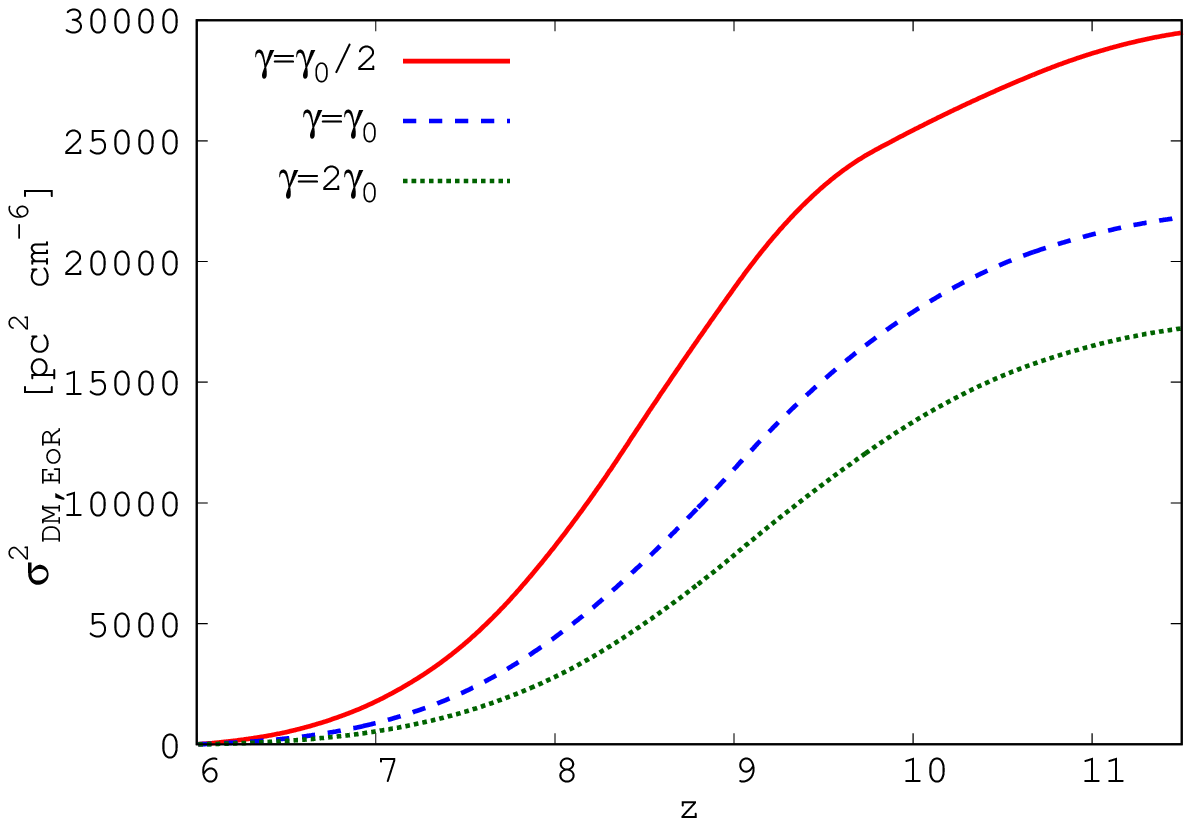}
\caption{The redshift evolution of $\sigma^2_{\rm DM, EoR}$ for models with different values of $\sigma_{\rm L}^2$ and $\gamma$. In the top panel, solid, dashed and doted lines have $2 \sigma_{\rm L0}^2$, $\sigma_{\rm L0}^2$ and $\sigma_{\rm L0}^2 / 2$, respectively. In the bottom panel, solid, dashed and doted lines have $2 \gamma_0$, $\gamma_0$ and $\gamma_0 / 2$. We assume fiducial BSD with ($\sigma_{\rm L0}^2$, $\gamma_0$) = (0.3, -1.7). } 
\label{fig:vari} 
\end{figure}

Let us see the effect of the functional form of the BSD, which is characterized by the variance and skewness. In the top panel of Fig.~\ref{fig:vari}, we compare models with $\sigma_{\rm L}^2 = 2 \sigma_{\rm L0}^2$, $\sigma_{\rm L0}^2$ and $\sigma_{\rm L0}^2 / 2$, where $\sigma_{\rm L}^2 = 0.3$. We find that larger variance leads to a larger value of $\sigma_{\rm DM,EoR}^2$. This is because a larger maximum bubble size is expected for a model with a larger value of variance and shifts the peak in the power spectrum to a smaller wave length (a larger scale). In addition, in the bottom panel of Fig.~\ref{fig:vari}, we compare the models with $\gamma = 2 \gamma_0$, $\gamma_0$ and $\gamma_0 / 2$, where $\gamma = -1.7$. The dependence on $\gamma$ can also be understood in terms of the maximum bubble size. For a fixed mean size, the maximum size is expected to be smaller for more skewed (larger $|\gamma|$) BSD.

\begin{figure}
\centering
\includegraphics[width=6.5cm]{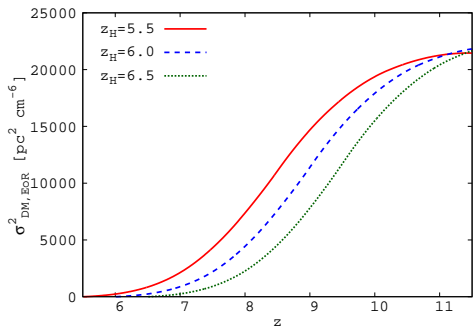}
\caption{The redshift evolution of $\sigma^2_{\rm DM, EoR}$ for the models with $z_{\rm H}=$ 5.5, 6.0 and 6.5. } 
\label{sigmazre} 
\end{figure}

Finally, we study the effect of ionization history on the $\sigma_{\rm DM,\,EoR}$. We compare three ionization models with different redshifts of reionization completion, $z_{\rm H} =$ 5.5, 6.0 and 6.5, as shown in Fig.~\ref{DM}. The evolution of $\sigma_{\rm DM}$ is plotted in Fig~\ref{sigmazre}. As one can see, the change of $z_{\rm H}$ just shifts the evolution of $\sigma_{\rm DM,\,EoR}$.  However, it should be noted that the observable $\sigma_{\rm DM}$ is contributed from IGM after reionization, while Fig.~\ref{sigmazre} shows a contribution only during EoR. Therefore, the model with $z_{\rm H} = 6.5$ has larger observable $\sigma_{\rm DM}$ than other models at $z>11$.

\section{Detectability}\label{sec:Det}

To probe the bubble properties via $\sigma_{\rm DM,EoR}$, we need to extract it from the total DM variance. Because the total DM variance is expressed as an integration from $z = 0$ to the redshift of interest, the $\sigma^2_{\rm DM,EoR}$ is estimated as $\sigma_{\rm DM,EoR}^2(z) = \sigma_{\rm DM}^2(z) - \sigma_{\rm DM}^2(z_{\rm H})$. Thus, the DMs should be measured for the objects at the end of reionization and its variance must be estimated.

Here let us mention the DM evolution at low redshifts. \cite{2014ApJ...780L..33M} studied the mean and variance of DMs upto $z \approx 1$ considering several models of baryon distribution. It was found that the mean and variance grow rapidly from $z = 0$ to $z \sim 1$ and stay almost constant at $z \gtrsim 1$. The variance at $z \gtrsim 1$ depends on the model of baryon distribution ranging from 30,000 to 160,000. Below, we take $100,000$ as a typical value.

Next, we estimate the number of FRBs necessary to obtain a useful constraint on the typical size of ionized bubbles at a single redshift. Denoting the statistical errors in $\sigma^2_{\rm DM}(z)$ and $\sigma^2_{\rm DM,EOR}(z)$ as $\delta(z)$ and $\delta_{\rm EOR}(z)$, respectively, we have $\delta_{\rm EOR}(z) = \sqrt{(\delta(z))^2 + (\delta(z_{\rm H}))^2}$. When we have $N$ DM samples with $N \gg 1$, $\delta(z)$ is evaluated as $\sqrt{2/N} \sigma^2_{\rm DM}(z)$. If we assume the number of samples at $z_{\rm H}$ is much larger than that at the redshift of interest ($z > z_{\rm H}$), $\delta(z_{\rm H}$) can be neglected and we have $\delta_{\rm EOR}(z) \sim \delta(z)$. Considering a nominal redshift $z = 9$, the typical values of $\sigma^2_{\rm DM}(z)$ and $\sigma^2_{\rm DM,EOR}(z)$ are $100,000$ and $11,000$ (see Fig.~\ref{sigma}), respectively. To distinguish our fiducial model and a model with $R_{\rm c} = 2R_{\rm c0}$, we need an accuracy of about $8,000$ so that the minimum necessary number of samples is estimated to be 400. In the same way, we evaluate the number of samples required for distinguishing the models with $R_{\rm c}$ and $2R_{\rm c}$ at four redshifts. We tabulate them in Table~\ref{table1} and show the errors as vertical lines in Fig.~\ref{sigma}. 

\begin{table}
  \begin{center}
    \caption{Required number of samples to distinguish models with $R_{\rm c}$ and $2R_{\rm c}$. }
    \begin{tabular}{|l|c|r||r||r|} \hline
      z & 8 & 9 &10&11 \\ \hline
      $\delta_{\rm EoR}\times10^{-3}$ & 3.9 & 8.0& 82& 75 \\ \hline
      N& 1450 & 390 & 420& 530 \\ \hline
    \end{tabular}
  \end{center}
  \label{table1}
\end{table} 

Here we ignored the DMs contributed from the Milky Way and the host galaxy of FRBs. According to \cite{2015MNRAS.451.4277D}, the disc and spiral arm of the Milky Way contribute to the DM by less than few hundreds $\rm pc\, cm^{-3}$ except galactic center and the contribution from the galactic halo is about $30~{\rm pc \,cm^{-3}}$. Thus, they are expected to be negligible for our purpose. 

The existence of FRBs during EoR is of critical importance in our method and it is deeply related to the origin of FRBs. Although the origin itself is outside the scope of this paper, let us mention the possibility of detecting such FRBs with the SKA in future. So far, while the FRB reported in \cite{2016MNRAS.460L..30C} has a large DM$\sim$1600 indicating it is located at $z>1$, we have no evidence of the existence of FRBs at such high redshifts ($z \gtrsim 6$). Nevertheless, in \cite{2016JCAP...05..004F}, they showed that, assuming the observation time of 100~ms for SKA\_{}LOW and 1ms for SKA\_{}MID, FRBs at $z=10$ with a peak luminosity as large as $5\times10^{33}[\rm erg\,s^{-1}\,Hz^{-1}\,sr^{-1}]$ found in \cite{2015Natur.528..523M} can be detected by SKA1. {Besides, the high detection rate is also crucial in our method. Recently,  \cite{2016MNRAS.460L..30C,2016arXiv161100458L,2016MNRAS.455.2207R} estimated the detection rate of FRBs with flux density 1 Jy ms to be $2 \times 10^3$ per sky per day at $z<1$. Based on the estimation,} \cite{2017arXiv170606582F} found {the detection rate, $6000\,[\rm sky^{-1}\,day^{-1}]$ from the EoR, as observed by SKA2\_{}MID.} Thus, considering that FRB search with the SKA will be performed as a commensal observation with pulsar search, there is a possibility for the detection of a relatively large number of FRBs at EoR, required by our method.

\section{Summary and Discussion}\label{sec:Con}

In this paper, we studied the effect of ionized bubbles on the DM variance of objects during the Epoch of Reionization. The mean DM and the variance were calculated by using a simple model of spherical ionized bubbles with a reasonable size distribution implied by more sophisticated models. We found that the DM variance is sensitive to the probability distribution of bubble size and can be used to probe it. Then the statistical error of the variance was evaluated and the number of sources necessary to obtain useful information is estimated to be about several hundreds to 1,000 at a single redshift. Because the bubble size reflects the abundance of ionizing sources and their emissivity, the DM variance can be a powerful tool to probe the type of ionizing source. Even if only a very small fraction of observed FRBs are located at high redshifts, detection of several hundreds of FRBs would be possible with an unprecedented survey of the SKA, although follow-up observations to determine the redshift of the host galaxy is necessary.

We focused on objects which emit UV photons such as star forming galaxies. They are regarded as a primary candidate of ionizing sources during the EoR. However, if quasars are main sources, the distribution of free electrons would be very different from the one considered here, because quasars emit X-rays which have much larger mean free paths compared to UV photons. Then, the power spectrum is larger at large scales since the typical bubble size is enhanced \citep{2017arXiv170104408K} and, thus, the $\sigma_{\rm DM}$ becomes large.

As discussed above, the measurement of $\sigma_{\rm DM,\,EoR}$ can probe the BSD and then we can interpret the result as the property of ionizing sources. For example, a small value of $\sigma_{\rm DM\,EoR}$ indicates that there are many ionizing sources with low ionizing efficiency which provide small bubbles. This suggests, for example, that weak feedback effect allows small halo to make stars and the escape fraction of heavy galaxy is low. Thus, the observation of $\sigma_{\rm DM}$ could become a powerful tool to probe these early objects and is complementary to 21cm-line observations.

\section*{Acknowledgement}

We are grateful to Dr. H. Shimabukuro for helpful comments on a draft version of this manuscript. This work is supported by Grant-in-Aid from the Ministry of Education, Culture, Sports, Science and Technology (MEXT) of Japan, Nos. 24340048, 26610048, 15H05896, 16H05999, 17H01110 (K.T.) and 16J01585 (S.Y.), and Bilateral Joint Research Projects of JSPS (K.T.).




\bsp	
\label{lastpage}

\begin{thebibliography}{}

\bibitem[Akahori et al.(2016)]{2016ApJ...824..105A} Akahori, T., Ryu, D., \& Gaensler, B.~M.\ 2016, \apj, 824, 105 

\bibitem[Burke-Spolaor \& Bannister(2014)]{2014ApJ...792...19B} Burke-Spolaor, S., \& Bannister, K.~W.\ 2014, \apj, 792, 19 

\bibitem[Bannister et al.(2017)]{2017ApJ...841L..12B} Bannister, K.~W., Shannon, R.~M., Macquart, J.-P., et al.\ 2017, \apjl, 841, L12 

\bibitem[Caleb et al.(2017)]{2017MNRAS.468.3746C} Caleb, M., Flynn, C., Bailes, M., et al.\ 2017, \mnras, 468, 3746 

\bibitem[Champion et al.(2016)]{2016MNRAS.460L..30C} Champion, D.~J., Petroff, E., Kramer, M., et al.\ 2016, MNRAS, 460, L30 

\bibitem[Dolag et al.(2015)]{2015MNRAS.451.4277D} Dolag, K., Gaensler, B.~M., Beck, A.~M., \& Beck, M.~C.\ 2015, MNRAS, 451, 4277 

\bibitem[Fan et al.(2006)]{2006AJ....132..117F} Fan, X., Strauss, M.~A., Becker, R.~H., et al.\ 2006, AJ, 132, 117 

\bibitem[Fialkov \& Loeb(2016)]{2016JCAP...05..004F} Fialkov, A., \& Loeb, A.\ 2016, J. Cosmology Astropart. Phys., 5, 004 

\bibitem[Fialkov \& Loeb(2017)]{2017arXiv170606582F} Fialkov, A., \& Loeb, A.\ 2017, arXiv:1706.06582

\bibitem[Furlanetto et al.(2004)]{2004ApJ...613....1F} Furlanetto, S.~R., 
Zaldarriaga, M., \& Hernquist, L.\ 2004, \apj, 613, 1 

\bibitem[Furlanetto \& Oh(2005)]{2005MNRAS.363.1031F} Furlanetto, S.~R., \& Oh, S.~P.\ 2005, \mnras, 363, 1031 

\bibitem[Furlanetto et al.(2006)]{2006MNRAS.365..115F} Furlanetto, S.~R., McQuinn, M., \& Hernquist, L.\ 2006, \mnras, 365, 115 

\bibitem[Friedrich et al.(2011)]{2011MNRAS.413.1353F} Friedrich, M.~M., 
Mellema, G., Alvarez, M.~A., Shapiro, P.~R., 
\& Iliev, I.~T.\ 2011, MNRAS, 413, 1353 

\bibitem[Giri et al.(2017)]{2017arXiv170600665G} Giri, S.~K., Mellema, G., Dixon, K.~L., \& Iliev, I.~T.\ 2017, arXiv:1706.00665 

\bibitem[Gleser et al.(2006)]{2006MNRAS.370.1329G} Gleser, L., Nusser, A., Ciardi, B., \& Desjacques, V.\ 2006, \mnras, 370, 1329 

\bibitem[Hasegawa \& Semelin(2013)]{2013MNRAS.428..154H} Hasegawa, K., \& Semelin, B.\ 2013, MNRAS, 428, 154 

\bibitem[Hong et al.(2014)]{2014JKAS...47...49H} Hong, S.~E., Ahn, K., Park, C., et al.\ 2014, Journal of Korean Astronomical Society, 47, 49 

\bibitem[Iliev et al.(2006)]{2006MNRAS.369.1625I} Iliev, I.~T., Mellema, G., Pen, U.-L., et al.\ 2006, \mnras, 369, 1625

\bibitem[Inoue(2004)]{2004MNRAS.348..999I} Inoue, S.\ 2004, MNRAS, 348, 999 

\bibitem[Ioka(2003)]{2003ApJ...598L..79I} Ioka, K.\ 2003, ApJ, 598, L79 

\bibitem[Iwata et al.(2009)]{Iwata09} Iwata., et al.\ 2009, \apj, 692, 1287

\bibitem[Johnston et al.(2017)]{2017MNRAS.465.2143J} Johnston, S., Keane, E.~F., Bhandari, S., et al.\ 2017, MNRAS, 465, 2143 

\bibitem[Kakiichi et al.(2017)]{2017arXiv170202520K} Kakiichi, K., Majumdar, S., Mellema, G., et al.\ 2017, arXiv:1702.02520

\bibitem[Keane et al.(2011)]{2011MNRAS.415.3065K} Keane, E.~F., Kramer, M., Lyne, A.~G., Stappers, B.~W., \& McLaughlin, M.~A.\ 2011, \mnras, 415, 3065 

\bibitem[Keane et al.(2012)]{2012MNRAS.425L..71K} Keane, E.~F., Stappers, B.~W., Kramer, M., \& Lyne, A.~G.\ 2012, MNRAS, 425, L71 

\bibitem[Keane et al.(2016)]{2016Natur.530..453K} Keane, E.~F., Johnston, S., Bhandari, S., et al.\ 2016, \nat, 530, 453 

\bibitem[Kulkarni et al.(2017)]{2017arXiv170104408K} Kulkarni, G., Choudhury, T.~R., Puchwein, E., \& Haehnelt, M.~G.\ 2017, arXiv:1701.04408 

\bibitem[Lawrence et al.(2016)]{2016arXiv161100458L} Lawrence, E., Vander Wiel, S., Law, C.~J., Burke Spolaor, S., \& Bower, G.~C.\ 2016, arXiv:1611.00458 

\bibitem[Lorimer et al.(2007)]{2007Sci...318..777L} Lorimer, D.~R., Bailes, M., McLaughlin, M.~A., Narkevic, D.~J., \& Crawford, F.\ 2007, Science, 318, 777 

\bibitem[Lorimer et al.(2013)]{2013MNRAS.436L...5L} Lorimer, D.~R., Karastergiou, A., McLaughlin, M.~A., \& Johnston, S.\ 2013, MNRAS, 436, L5 

\bibitem[Masui et al.(2015)]{2015Natur.528..523M} Masui, K., Lin, H.-H., Sievers, J., et al.\ 2015, \nat, 528, 523 

\bibitem[McGreer et al.(2015)]{2015MNRAS.447..499M} McGreer, I.~D., Mesinger, A., \& D'Odorico, V.\ 2015, \mnras, 447, 499 

\bibitem[McQuinn(2014)]{2014ApJ...780L..33M} McQuinn, M.\ 2014, ApJ, 780, L33 

\bibitem[Mesinger et al.(2011)]{2011MNRAS.411..955M} Mesinger, A., Furlanetto, S., \& Cen, R.\ 2011, \mnras, 411, 955 

\bibitem[Mesinger et al.(2016)]{2016MNRAS.459.2342M} Mesinger, A., Greig, B., \& Sobacchi, E.\ 2016, MNRAS, 459, 2342 

\bibitem[Okamoto et al.(2008)]{2008MNRAS.390..920O} Okamoto, T., Gao, L., \& Theuns, T.\ 2008, MNRAS, 390, 920 

\bibitem[Petroff et al.(2015)]{2015MNRAS.447..246P} Petroff, E., Bailes, M., Barr, E.~D., et al.\ 2015, \mnras, 447, 246 

\bibitem[Petroff et al.(2017)]{2017MNRAS.469.4465P} Petroff, E., Burke-Spolaor, S., Keane, E.~F., et al.\ 2017, \mnras, 469, 4465 

\bibitem[Planck Collaboration et al.(2015)]{2015arXiv150201589P} Planck 
Collaboration, Ade, P.~A.~R., Aghanim, N., et al.\ 2015, arXiv:1502.01589 

\bibitem[Planck Collaboration et al.(2016)]{2016A&A...596A.108P} Planck Collaboration, Adam, R., Aghanim, N., et al.\ 2016, \aap, 596, A108 

\bibitem[Pritchard \& Furlanetto(2007)]{2007MNRAS.376.1680P}
Pritchard, J.~R., \& Furlanetto, S.~R.\ 2007, MNRAS, 376, 1680

\bibitem[Ravi et al.(2015)]{2015ApJ...799L...5R} Ravi, V., Shannon, R.~M., \& Jameson, A.\ 2015, \apjl, 799, L5 

\bibitem[Rane et al.(2016)]{2016MNRAS.455.2207R} Rane, A., Lorimer, D.~R., Bates, S.~D., et al.\ 2016, \mnras, 455, 2207 

\bibitem[Shimabukuro et al.(2015)]{2015MNRAS.451..467S} Shimabukuro, H., Yoshiura, S., Takahashi, K., Yokoyama, S., \& Ichiki, K.\ 2015, MNRAS, 451, 467 

\bibitem[Shimabukuro et al.(2017)]{2017MNRAS.468.1542S} Shimabukuro, H., Yoshiura, S., Takahashi, K., Yokoyama, S., \& Ichiki, K.\ 2017, \mnras, 468, 1542 

\bibitem[Spitler et al.(2014)]{2014ApJ...790..101S} Spitler, L.~G., Cordes, J.~M., Hessels, J.~W.~T., et al.\ 2014, \apj, 790, 101 

\bibitem[Spitler et al.(2016)]{2016Natur.531..202S} Spitler, L.~G., Scholz, P., Hessels, J.~W.~T., et al.\ 2016, \nat, 531, 202 

\bibitem[Steidel et al.(2001)]{Steidel01} Steidel, C.~C., Pettini, M., Adelberger, K.~L.\ 2001, \apj, 546, 665

\bibitem[Tendulkar et al.(2017)]{2017arXiv170101100T} Tendulkar, S.~P., Bassa, C., Cordes, J.~M., et al.\ 2017, arXiv:1701.01100 

\bibitem[Thornton et al.(2013)]{2013Sci...341...53T} Thornton, D., Stappers, B., Bailes, M., et al.\ 2013, Science, 341, 53 

\bibitem[Totani(2013)]{2013PASJ...65L..12T} Totani, T.\ 2013, PASJ, 65,  

\bibitem[Yoshiura et al.(2017)]{2017MNRAS.465..394Y} Yoshiura, S., Shimabukuro, H., Takahashi, K., \& Matsubara, T.\ 2017, MNRAS, 465, 394

\bibitem[Zahn et al.(2007)]{2007ApJ...654...12Z} Zahn, O., Lidz, A., McQuinn, M., et al.\ 2007, \apj, 654, 12 

\bibitem[Zahn et al.(2011)]{2011MNRAS.414..727Z} Zahn, O., Mesinger, A., McQuinn, M., et al.\ 2011, MNRAS, 414, 727 

\bibitem[Yang \& Zhang(2016)]{2016ApJ...830L..31Y} Yang, Y.-P., \& Zhang, B.\ 2016, \apjl, 830, L31 

\bibitem[Zheng et al.(2014)]{2014ApJ...797...71Z} Zheng, Z., Ofek, E.~O., Kulkarni, S.~R., Neill, J.~D., \& Juric, M.\ 2014, \apj, 797, 71 

\bibitem[Zhou et al.(2014)]{2014PhRvD..89j7303Z} Zhou, B., Li, X., Wang, T., Fan, Y.-Z., \& Wei, D.-M.\ 2014, \prd, 89, 107303 


\end{thebibliography}
\end{document}